\documentclass[12pt]{iopart}

\newread\testifexists
\def\GetIfExists #1 {\immediate\openin\testifexists=#1
    \ifeof\testifexists\immediate\closein\testifexists\else
    \immediate\closein\testifexists\input #1\fi}

\usepackage{amsfonts}\usepackage{amssymb}\usepackage{url} \mathsurround=2pt
\immediate\openin\testifexists=e
    \ifeof\testifexists\immediate\closein\testifexists\else
    \immediate\closein\testifexists\input e\fipsf
\def\Bbb#1{\setbox0=\hbox{$\tt #1$}  \copy0\kern-\wd0\kern .1em\copy0}
\def\bbf#1{\setbox0=\hbox{$#1$} \kern-.025em\copy0\kern-\wd0
        \kern.05em\copy0\kern-\wd0 \kern-.025em\raise.0433em\box0}
\def\a{\alpha}      \def\b{\beta}         
\def\d{\delta}        \def\e{\varepsilon}
\def\h{\eta}          \def\l{\lambda}     \def\L{\Lambda}
\def\m{\mu}             \def\F{\Phi}        
\def\n{\nu}         \def\j{\psi}    
       
\def\t{\tau}        \def\th{\theta}  
\def\x{\xi}              
\def\w{\omega}        

 \def\LL{{\cal L}} 
\def\pa{\partial} \def\ra{\rightarrow}
\def\na{\nabla}
 
\def\dd{{\rm d}}     \def\ket{\rangle}

\def\deff{\ {\buildrel{\rm def}\over{=}}\ }

\def\fract#1#2{{\textstyle{#1\over#2}}}
\def\ffract#1#2{\raise .3 em\hbox{$\scriptstyle#1$}\kern-.25em/
                \kern-.2em\lower .2 em \hbox{$\scriptstyle#2$}}

\def\half{\fract12}  

\def\part#1#2{{\partial#1\over\partial#2}}

\newcommand{\be}{\begin{eqnarray}}
\newcommand{\ee}{\end{eqnarray}}
\newcommand{\eqn}[1]{(\ref{#1})}
\newcommand{\nn}{\nonumber\\}

\newcommand{\itm}[1]{\item[#1]}

\newcommand{\newsec}[1]{\section{#1}\setcounter{equation}{0}}
\def\printversion{\setlength{\textheight}{9in}\setlength{\oddsidemargin}{0in}
    \setlength{\textwidth}{6.3in}\setlength{\topmargin}{-0.1in}}

 \newcommand {\eel}[1]{\label{#1}\end{eqnarray}} \newcommand{\crl}[1]{\label{#1}\\} % equationnumbers

 \printversion

\begin{document}
\begin{flushright}
{\small
ITP-UU-06/06\\
SPIN-06/04\\
gr-qc/0602076\\
}
\end{flushright}
\title[Invariance under complex transformations]{Invariance under complex transformations, and its relevance to the cosmological constant problem}
\author{Gerard 't~Hooft, Stefan Nobbenhuis}
\address{Institute for Theoretical Physics \\
Utrecht University, Leuvenlaan 4\\ 3584 CC Utrecht, the
Netherlands\medskip \\ and
\medskip \\ Spinoza Institute \\ Postbox 80.195 \\ 3508 TD
Utrecht, the Netherlands \smallskip}
\eads{\mailto{g.thooft@phys.uu.nl},\,\,\,\mailto{S.J.B.Nobbenhuis@phys.uu.nl}}
\begin{abstract}
In this paper we study a new symmetry argument that results in a
vacuum state with strictly vanishing vacuum energy. This argument
exploits the well-known feature that de Sitter and Anti- de Sitter
space are related by analytic continuation. When we drop boundary
and hermiticity conditions on quantum fields, we get as many
negative as positive energy states, which are related by
transformations to complex space. The paper does not directly
solve the cosmological constant problem, but explores a new
direction that appears worthwhile.
\end{abstract}

\submitto{Classical \& Quantum Gravity}

\maketitle

\newsec{Introduction}\label{intro.sec}

The cosmological constant problem is one of the major obstacles for both particle physics and cosmology. The question is why is the effective
cosmological constant, $\Lambda_{eff}$, defined as $\Lambda_{eff} = \Lambda + 8\pi G\langle\rho\rangle$ so close to zero\footnote{Note that using
this definition we use units in which the cosmological constant has dimension $\mbox{GeV}^2$ throughout.}. As is well known, different contributions
to the vacuum energy density from particle physics would naively give a value for $\langle\rho\rangle$ of order $M_P^4$, which then would have to be
(nearly) cancelled by the unknown `bare' value of $\Lambda$.

This cancellation has to be precise to about 120~decimal places if we compare the zero-point energy of a scalar field, using the Planck scale as a
cutoff, and the experimental value of $\rho_{vac}=\langle\rho\rangle + \Lambda/8\pi G$, being $10^{-47} \mbox{GeV}^4$. As is well known, even if we
take a TeV scale cutoff the difference between experimental and theoretical results still requires a fine-tuning of about 50 orders of magnitude.
This magnificent fine-tuning seems to suggest that we fail to observe something that is absolutely essential. In a recent paper, one of
us\cite{Nobbenhuis:2004wn}gave a categorization of the different proposals that have occurred in the literature and pointed out for each of them
where the shortcomings are.

In this paper we discuss in more detail a scenario that has been introduced in \cite{Nobbenhuis:2004wn}, based on symmetry with respect to a
transformation towards imaginary values of the space-time coordinates: \(x^\m\ra i\,x^\m\). This symmetry entails a new definition of the vacuum
state, as the unique state that is invariant under this transformation. Since curvature switches sign, this vacuum state must be associated with zero
curvature, hence zero cosmological constant. The most striking and unusual feature of the symmetry is the fact that the \emph{boundary conditions} of
physical states are not invariant. Physical states obey boundary conditions when the real parts of the coordinates tend to infinity, not the
imaginary parts. This is why all physical states, except the vacuum, must break the symmetry. We will argue that a vanishing cosmological constant
could be a consequence of the specific boundary conditions of the vacuum, upon postulating this complex symmetry.

We do not address the issue of non-zero cosmological constant, nor the so-called cosmic coincidence problem. We believe that a symmetry which would
set the cosmological constant to exactly zero would be great progress.

The fact that we are transforming real coordinates into imaginary coordinates implies, \textit{inter alia}, that hermitean operators are transformed
into operators whose hermiticity properties are modified. Taking the hermitean conjugate of an operator requires knowledge of the boundary conditions
of a state. The transition from \(x\) to \(ix\) requires that the boundary conditions of the states are modified. For instance, wave functions \(\F\)
that are periodic in real space, are now replaced by waves that are exponential expressions of \(x\), thus periodic in \(ix\). But we are forced to
do more than that. Also the creation and annihilation operators will transform, and their commutator algebra in complex space is not a priori clear;
it requires careful study.

Thus, the symmetry that we are trying to identify is a symmetry of laws of nature \emph{prior} to imposing any boundary conditions. Demanding
invariance under \(x_\m\ra x_\m + a_\m\) where \(a_\m\) may be real or imaginary, violates boundary conditions at \(\F\ra\infty\), leaving only one
state invariant: the \emph{physical} vacuum.

\newsec{Classical Scalar Field}\label{classfield.sec}

To set our notation, consider a real, classical, scalar field
\(\F(x)\) in \(D\) space-time dimensions, with Lagrangian \be
\LL=-\half(\pa_\m\F)^2-V(\F(x))\ ,\qquad V(\F)=\half
m^2\F^2+\l\F^4\ . \eel{Lclass} Adopting the metric convention
\((-+++)\), we write the energy-momentum tensor as \be
T_{\m\n}(x)=\pa_\m\F(x)\pa_\n\F(x)+g_{\m\n}\LL(\F(x))\ .
\eel{Tmunuclass} The Hamiltonian \(H\) is \be \fl\quad
H=\int\dd^{D-1}\vec x\,T_{00}(x)\ ;\qquad T_{00}=
\half\Pi^2+\half(\vec\pa\F)^2+V(\F)\ ;\qquad\Pi(x)=\pa_0\F(x)\ .
\eel{Hamclass}

Write our transformation as \(x^\m=iy^\m\), after which all coordinates are rotated in their complex planes such that \(y^\m\) will become real. For
redefined notions in \(y\) space, we use subscripts or superscripts \(y\), e.g., \(\pa_\m^y=i\pa_\m\). The field in \(y\) space obeys the Lagrange
equations with \be
\LL_y&=&-\LL\ =\ -\half(\pa^y_\m\F)^2+V(\F)\ ;\\
T_{\m\n}^y&=&-T_{\m\n}\ =\ \pa_\m^y\F(iy)\pa^y_\n\F(iy)+g_{\m\n}\LL_y(\F(iy))\ .
\eel{LTy} The Hamiltonian in \(y\)-space is \be H=-(i^{D-1})H_y&,&\quad
H_y=\int\dd^{D-1}y\,T_{00}^y\ ;\\T_{00}^y=\half\Pi_y^2+\half(\vec\pa_y\F)^2-V(\F)
&,&\quad\Pi_y(y)=i\Pi(iy)\ .\eel{Tyclass}

If we keep only the mass term in the potential, \(V(\F)=\half m^2\F^2\), the field obeys the Klein-Gordon equation. In the real \(x\)-space, its
solutions can be written as \be &&\F(x,\,t)\ =\int \dd^{D-1} p \left( a(p)e^{i(px)}+a^\ast(p) e^{-i(px)}\right)\ ,\crl{fieldfourrier} &&\Pi(x,\,t)\
=\int \dd^{D-1} p\ p^0\left(-ia(p)e^{i(px)}+ia^\ast(p) e^{-i(px)}\right)\ ;\crl{Pifourrier} &&p^0=\sqrt{\vec p^{\;2}+m^2}\ ,\qquad(px)\deff \vec
p\cdot\vec x-p^0t\ , \eel{pnulclass} where \(a(p)\) is just a c-number.

Analytically continuing these solutions to complex space, yields:
\be \fl\quad\F(iy,\,i\t)&=&\int \dd^{D-1}q \left(
a_y(q)e^{i(qy)}+\hat{a_y}(q) e^{-i(qy)}\right)\
,\crl{fieldyfourrier} \fl\quad \Pi_y(y,\,\t)\ =\
i\Pi(iy,\,i\t)&=&\int
\dd^{D-1}q\,q^0\left(-ia_y(q)e^{i(qy)}+i\hat{a_y}(q)
e^{-i(qy)}\right)\ ;\qquad{}\crl{Piyfourrier}\fl\quad
&&q^0=\sqrt{\vec q^{\;2}-m^2}\ ,\qquad(qy)\deff \vec q\cdot\vec
y-q^0\t\ . \eel{qnul} The new coefficients could be analytic
continuations of the old ones, \be a_y(q)= (-i)^{D-1}a(p)\ ,\qquad
\hat{a_y}(q)= (-i)^{D-1}a^\ast(q)\ ,\qquad p^\m=-iq^\m\ ,\eel{aaq}
but this makes sense only if the \(a(p)\) would not have
singularities that we cross when shifting the integration contour.
Note, that, since \(D=4\) is even, the hermiticity relation
between \(a_y(q)\) and \({\hat a_y}(q)\) is lost. We can now
consider solutions where we restore them: \be {\hat
a_y}(q)=a_y^\ast(q)\ , \eel{hatstar} while also demanding
convergence of the \(q\) integration. Such solutions would not
obey acceptable boundary conditions in \(x\)-space, and the fields
would be imaginary rather than real, so these are unphysical
solutions. The important property that we concentrate on now,
however, is that, according to Eq.~\eqn{LTy}, these solutions
would have the opposite sign for \(T_{\m\n}\).

Of course, the field in \(y\)-space appears to be tachyonic, since
\(m^2\) is negative. In most of our discussions we should put
\(m=0\). A related transformation with the objective of
\(T_{\m\n}\ra -T_{\m\n}\) was made by Kaplan and Sundrum in
\cite{Kaplan:2005rr}. Non-Hermitian Hamiltonians were also studied
by Bender et al. in for example
\cite{Bender:2005pf,Bender:2005hf,Bender:1998ke,Bender2005}.
Another approach based on similar ideas which tries to forbid a
cosmological constant can be found in \cite{Bonelli:2000tz}.

\newsec{Gravity}

Consider Einstein's equations: \be R_{\m\n} - \half g_{\m\n}R - \L g_{\m\n} = -8\pi GT_{\m\n}. \ee Writing \be x^\m=i\,y^\m=i(\vec y,\,\t)\ , \qquad
g_{\m\n}^y(y)\ra g_{\m\n}(x=iy), \eel{compltrf} and defining the Riemann tensor in \(y\) space using the derivatives \(\pa^y_\m\), we see that \be
R_{\m\n}^y=-R_{\m\n}(iy)\ . \eel{Ricci} Clearly, in \(y\)-space, we have the equation \be R_{\m\n}^y - \half g_{\m\n}^yR^y + \L g_{\m\n}^y = +8\pi
GT_{\m\n}(iy)= -8\pi GT_{\m\n}^y. \ee Thus, Einstein's equations are invariant except for the cosmological constant term.

A related suggestion was made in \cite{Erdem2004}. In fact, we could consider formulating the equations of nature in the full complex space
\(z=x+iy\), but then everything becomes complex. The above transformation is a one-to-one map from real space \(\Re^3\) to the purely imaginary space
\(\Im^3\), where again real equations emerge.

The transformation from real to imaginary coordinates naturally relates deSitter space with anti-deSitter space, or, a vacuum solution with positive
cosmological constant to a vacuum solution with negative cosmological constant. Only if the cosmological constant is zero, a solution can map into
itself by such a transformation. None of the excited states can have this invariance, because they have to obey boundary conditions, either in real
space, or in imaginary space.

\newsec{Non-relativistic Particle} The question is now, how much of this survives in a quantum theory. The simplest example to be discussed is the
non-relativistic particle in one space dimension. Consider the Hamiltonian \be H={p^2\over 2m}+V(x)\,,\eel{NRHamilton} where \(p=-i\pa/\pa x\).
Suppose that the function \(V(x)\) obeys \be V(x)=-V(ix)\ ,\qquad V(x)=x^2V_0(x^4)\ , \eel{Potentl} with, for instance, \(V_0(x^4)=e^{-\l x^4}\).
Consider a wave function \(|\j(x)\ket\) obeying the wave equation \(H|\j\ket=E|\j\ket\). Then the substitution \be x=iy\ ,\qquad p=-ip_y\ ,\qquad
p_y=-i{\pa\over\pa y}\ ,\eel{ComplTransf} gives us a new function \(|\j(y)\ket\) obeying \be H_y|\j(y)\ket=-E|\j(y)\ket\ ,\qquad
H_y={p_y^2\over2m}+V(y)\ .\eel{NewEq} Thus, we have here a symmetry transformation sending the hamiltonian \(H\) into \(-H\). Clearly, \(|\j(y)\ket\)
cannot in general be an acceptable solution to the usual Hamilton eigenvalue equation, since \(|\j(y)\ket\) will not obey the boundary condition
\(|\j(y)|^2\ra 0\) if \(y\ra \pm\infty\). Indeed, hermiticity, normalization, and boundary conditions will not transform as in usual symmetry
transformations.

Yet, this symmetry is not totally void. If \(V=0\), a state
\(|\j_0\ket\) can be found that obeys both the boundary conditions
at \(x\ra\pm\infty\) and \(y\ra\pm\infty\). It is the ground
state, \(\j(x)=\mathrm{constant}\). It obeys both boundary
conditions because of its invariance under transformations \(x\ra
x+a\), where \(a\) can be any complex number. Because of our
symmetry property, it obeys \(E=-E\), so the energy of this state
has to vanish. Since it is the only state with this property, it
must be the ground state. Thus, we see that our complex symmetry
may provide for a mechanism that generates a zero-energy ground
state, of the kind that we are looking for in connection with the
cosmological constant problem.

In general, if \(V(x)\ne0\), this argument fails. The reason is
that the invariance under complex translations breaks down, so
that no state can be constructed obeying all boundary conditions
in the complex plane. In our treatment of the cosmological
constant problem, we wish to understand the physical vacuum. It is
invariant under complex translations, so there is a possibility
that a procedure of this nature might apply.

As noted by Jackiw \cite{Jackiw}, there is a remarkable example in
which the potential does not have to vanish. We can allow for any
well-behaved function that depends only on \(x^4=y^4\). For
example, setting \(m=1\), \be V(x)=2x^6-3x^2=x^2(2x^4-3),\ee with
ground state wavefunction \(\exp(-x^4/2)\), indeed satisfies
condition \eqn{Potentl}, which guarantees zero energy eigenvalue.
Note that this restricts the transformation to be discrete, since
otherwise it crosses the point \(x=\sqrt{i}y\) where the potential
badly diverges. Boundary conditions are still obeyed on the real
and imaginary axis, but not for general complex values, see figure
\ref{xixfig3.fig}.

\begin{figure}[ht] \setcounter{figure}{0} \begin{quotation}
 \epsfxsize=62 mm\epsfbox{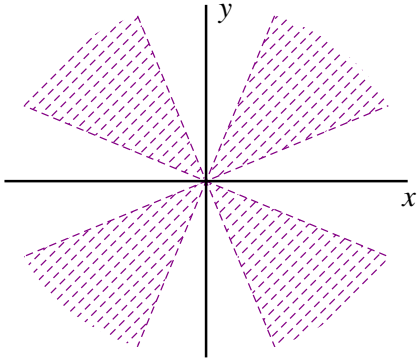}
  \caption{\footnotesize{Region in complex space where the potential is
  well-defined; the shaded region indicates where boundary conditions are not obeyed.}}
  \label{xixfig3.fig} \end{quotation}
\end{figure}

Moreover, as Jackiw also pointed out, this example is intriguing
since it reminds us of supersymmetry. Setting again \(m=1\) for
clarity of notation, the Hamiltonian \be
H=\frac{1}{2}(p+iW')(p-iW')\ee with \(W\) the superpotential and a
prime denoting a derivative with respect to the fields, has a
scalar potential \be V=\frac{1}{2}(W'W'-W'').\ee If \(W\)
satisfies condition \eqn{Potentl}, the Hamiltonian possesses a
zero energy eigenfunction \(e^{-W}\), which obeys the correct
boundary conditions in \(x\) and \(y\). The Hamiltonian in this
example is the bosonic portion of a supersymmetric Hamiltonian, so
our proposal might be somehow related to supersymmetry.

We need to know what happens with hermiticity and normalizations. Assume the usual hermiticity properties of the bras, kets and the various operators
in \(x\) space. How do these properties read in \(y\) space? We have \be x=x^\dagger&\quad&p=p^\dagger\ ,\nn y=-y^\dagger&\quad&p_y=-p_y^\dagger\
,\eel{hermt} but the commutator algebra is covariant under the transformation: \be [p,\,x] =-i&\quad&p=-i\pa/\pa x\ ,\nn \ [p_y,\,y]
=-i&\quad&p_y=-i\pa/\pa y\ . \eel{commt} Therefore, the wave equation remains the same\ locally in \(y\) as it is in \(x\), but the boundary
condition in \(y\) is different from the one in \(x\). If we would replace the hermiticity properties of \(y\) and \(p_y\) in Eq.~\eqn{hermt} by
those of \(x\) and \(p\), then we would get only states with \(E\le0\).

\newsec{Harmonic Oscillator}\label{harmonic.sec} An instructive example is the \(x\ra y\) transformation, with \(x=iy\), in the harmonic oscillator.
The Hamiltonian is \be H={p^2\over 2m}+\half m\w^2x^2\ ,\eel{harmham}for which one introduces the conventional annihilation and creation operators
\(a\) and \(a^\dagger\): \be &&a=\sqrt{m\w\over2}\left(x+{ip\over m\w}\right)\ ,\qquad a^\dagger=\sqrt{m\w\over2}\left(x-{ip\over m\w}\right)\ ;\\
&&H= \w(a^\dagger a+\half)\ . \eel{Hadaga} In terms of the operators in \(y\)-space, we can write \be a_y=\sqrt{m\w\over2}\left(y+{ip_y\over
m\w}\right)\ ,&\quad&
\hat{a_y}=\sqrt{m\w\over2}\left(y-{ip_y\over m\w}\right)\ =\ -a_y^\dagger\ ;\\
a_y=-ia^\dagger\ ,\quad \hat{a_y}=-i a\ ,&\quad& H= -\w(\hat{a_y} a_y+\half)\ . \eel{Hahatay}
If one were to replace the correct hermitian conjugate of \(a_y\) by \(\hat{a_y}\) instead of
\(-\hat{a_y}\), then the Hamiltonian \eqn{Hahatay} would take only the eigenvalues
\(H=-H_y=\w(-n-\half)\). Note that these form a natural continuation of the eigenstates
\(\w(n+\half)\), as if \(n\) were now allowed only to be a negative integer.

The ground state, \(|0\ket\) is not invariant. In \(x\)-space, the \(y\) ground state would be the non-normalizable state \(\exp(+\half m\w x^2)\),
which of course would obey `good' boundary conditions in \(y\)-space.

\newsec{Second Quantization}

\begin{figure}[ht] \begin{quotation}
 \epsfxsize=125 mm\epsfbox{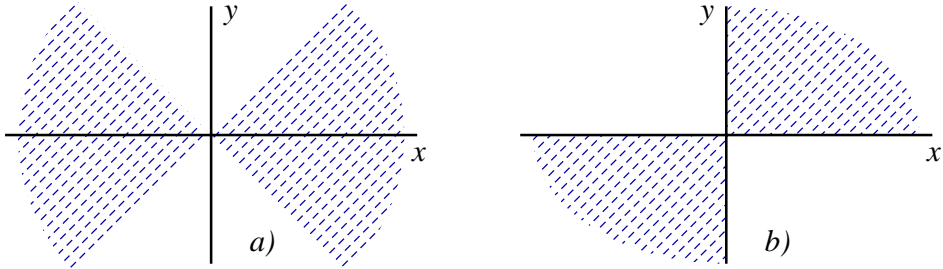}
  \caption{\footnotesize{Region in complex space where the Dirac delta function is
  well-defined, \((a)\) if \(\l\) is real, \((b)\) if \(\m\) is real and positive.}}
  \label{xixfig1.fig} \end{quotation}
\end{figure}

The examples of the previous two sections, however, are not the
transformations that are most relevant for the cosmological
constant. We wish to turn to imaginary coordinates, but not to
imaginary oscillators. We now turn our attention to
second-quantized particle theories, and we know that the vacuum
state will be invariant, at least under all complex translations.
Not only the hermiticity properties of field operators are
modified in the transformation, but now also the commutation rules
are affected. A scalar field \(\F(x)\) and its conjugate,
\(\Pi(x)\), often equal to \(\dot\F(x)\), normally obey the
commutation rules \be [\Pi(\vec x,t),\,\F({\vec
x\,}',\,t)]=-i\d^3(\vec x-{\vec x\,}')\ , \eel{fieldcomm} where
the Dirac deltafunction \(\d(x)\) may be regarded as \be
\d(x)=\textstyle{\sqrt{\l\over\pi}}\, e^{-\l x^2}\ ,
\eel{Diracdel} in the limit \(\l\uparrow\infty\). If \(\vec x\) is
replaced by \(i\vec y\), with \(\vec y\) real, then the
commutation rules are \be [\Pi(i\vec y,t),\,\F(i{\vec
y\,}',\,t)]=-i\d^3(i(\vec y-{\vec y\,}'))\ , \eel{complfieldcomm}
but, in Eq.~\eqn{Diracdel} we see two things happen:
\begin{itemize}\itm{\((i)\)} This delta function does not go to zero unless its
argument \(x\) lies in the right or left quadrant of
Fig.~\ref{xixfig1.fig}. Now, this can be cured if we add an
imaginary part to \(\l\), namely \(\l\ra -i\m\), with \(\m\) real.
Then the function \eqn{Diracdel} exists if \(x=r\,e^{i\th}\), with
\(0<\th<\half\pi\). But then, \itm{\((ii)\)} If \(x=iy\), the sign
of \(\m\) is important. If \(\m>0\), replacing \(x=iy\), the delta
function becomes \be
\d(iy)=\textstyle{\sqrt{-i\m\over\pi}}\,e^{-i\m y^2}\ra-i\d(y)\
,\eel{idel} which would be \(+i\d(y)\) had we chosen the other
sign for \(\m\).\end{itemize} We conclude that the sign of the
square root in Eq.~\eqn{Diracdel} is ambiguous.

There is another way to phrase this difficulty. The commutation rule \eqn{fieldcomm} suggests that either the field \(\F(\vec x,\,t)\) or \(\Pi(\vec
x,\,t)\) must be regarded as a distribution. Take the field \(\Pi\). Consider test functions \(f(\vec x)\), and write \be\Pi(f,\,t)\deff\int f(\vec
x)\,\Pi(\vec x,t)\,\dd^3\vec x\ ;\qquad [\Pi(f,\,t),\,\F(\vec x,\,t)]=-if(\vec x)\ . \eel{commdistr} As long as \(\vec x\) is real, the integration
contour in Eq.~\eqn{commdistr} is well-defined. If, however, we choose \(x=iy\), the contour must be taken to be in the complex plane, and if we only
wish to consider real \(y\), then the contour must be along the imaginary axis. This would be allowed if \(\Pi(\vec x,\,y)\) is holomorphic for
complex \(\vec x\), and the end points of the integration contour should not be modified.

 \begin{figure}[ht]
\begin{quotation}
 \epsfxsize=125 mm\epsfbox{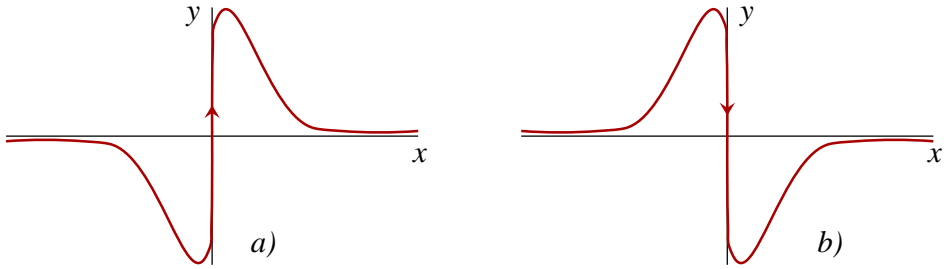}
  \caption{\footnotesize{Integration contour for the commutator algebra \eqn{commdistr},
\((a)\) and \((b)\) being two distinct choices.}}
  \label{xixfig2.fig}\end{quotation}
\end{figure}

For simplicity, let us take space to be one-dimensional. Assume that the contour becomes as in Fig.~\ref{xixfig2.fig}\textit{a}. In the \(y\) space,
we have \be \Pi(f,\,t) \deff \int_{-\infty}^\infty f(iy)\Pi(iy)\dd(iy)\ ;\qquad [\Pi(f,\,t),\,\F(i y,\,t)]=-if(i y)\ . \eel{compldistr} so that \be
[\Pi(iy,\,t),\,\F(iy',\,t)]=-\d(y-y')\ . \eel{complcomm} Note now that we could have chosen the contour of Fig.~\ref{xixfig2.fig}\textit{b} instead.
In that case, the integration goes in the opposite direction, and the commutator algebra in Eq.~\eqn{complcomm} receives the opposite sign. Note also
that, if we would be tempted to stick to one rule only, the commutator algebra would receive an overall minus sign if we apply the transformation
\(x\ra iy\) twice.

The general philosophy is now that, with these new commutation
relations in \(y\)-space, we could impose conventional hermiticity
properties in \(y\)-space, and then consider states as
representations of these operators. How do individual states then
transform from \(x\)-space to \(y\)-space or \textit{vice versa}?
We expect to obtain non-normalizable states, but the situation is
worse than that. Let us again consider one space-dimension, and
begin with defining the annihilation and creation operators \(a(
p)\) and \(a^\dagger(p)\) in \(x\)-space:

\be \F(x,\,t)=\quad\int{\dd p\over\sqrt{2\pi\cdot
2p^0}}\left(a(p)e^{i(px)}+a^\dagger(p)e^{-i(px)}\right)\ ,
\crl{fieldcreate} \Pi(x,\,t)=\ \ \int{\dd
p\sqrt{p^0}\over\sqrt{2\cdot2\pi}}\left(-ia(p)e^{i(px)}+ia^\dagger(p)e^{-i(px)}\right)\\\
\vphantom{\Big[}\qquad\qquad\qquad\qquad p^0=\sqrt{\vec
p^{\;2}+m^2}\ ,\qquad(px)\deff \vec p\cdot\vec x-p^0t\\
a(p)=\int{\dd
x\over\sqrt{2\pi\cdot2p^0}}\left(p^0\F(x,\,t)+i\Pi(x,\,t)\right)e^{-i(px)}\
,&&\crl{annihilfield} a^\dagger{(p)=\ \int{\dd
x\over\sqrt{2\pi\cdot2p^0}}\left(p^0\F(x,\,t)-i\Pi(x,\,t)\right)e^{i(px)}}\
.&&\ \eel{createfield}\\ Insisting that the commutation rules
\([a(p),\,a^\dagger(p')]=\d(p-p')\) should also be seen in
\(y\)-space operators: \be[a_y(q),\,\hat{a_y}(q')]=\d(q-q')\
,\eel{commycreate} we write, assuming \(p^0=-iq^0\) and
\(\Pi=-i\pa\F/\pa\t\) for free fields,

\be \F(iy,\,i\t)=\ \ \int{\dd q\over\sqrt{2\pi\cdot
2q^0}}\left(a_y(q)e^{i(qy)}+\hat{a_y}(q)e^{-i(qy)}\right)\
\crl{fieldycreate} \Pi(iy,\,i\t)=\qquad\int{\dd
q\sqrt{q^0}\over\sqrt{2\cdot2\pi}}\left(
-a_y(q)e^{i(qy)}+\hat{a_y}(q)e^{-i(qy)}\right)\ ,\crl{conjycreate}
\vphantom{\Big[}\qquad\qquad\qquad\qquad q^0=\sqrt{\vec
q^{\;2}-m^2}\ ,\qquad(qy)\deff \vec q\cdot\vec y-q^0\t\\
a_y(q)=\int{\dd y\over\sqrt{2\pi\cdot2q^0}}\left( q^0\F(iy,\,i\t)-
\Pi(iy,\,i\t)\right)e^{i(qy)}\ ,&&\crl{annihilyfield}
\hat{a_y}{(q)=\ \int{\dd y\over\sqrt{2\pi\cdot2q^0}}\left(
q^0\F(iy,\,i\t)+ \Pi(iy,\,i\t)\right)e^{-i(qy)}}\ ,&&\
\eel{createyfield}\\ so that the commutator \eqn{commycreate}
agrees with the field commutators \eqn{complcomm}. In most of our
considerations, we will have to take \(m=0\); we leave \(m\) in
our expressions just to show its sign switch.

In \(x\)-space, the fields \(\F\) and \(\pi\) are real, and the exponents in
Eqs~\eqn{fieldycreate}---\eqn{createyfield} are all real, so the hermiticity relations
are \(a_y^\dagger=a_y\) and \(\hat{a_y}^\dagger=\hat{a_y}\). As in the previous sections,
we replace this by \be \hat{a_y}=a_y^\dagger\ . \eel{aydagger}

The Hamiltonian for a free field reads \be
H&=&i\int_{-\infty}^\infty\dd
y\left(\half\Pi(iy)^2-\half(\pa_y\F(iy))^2+\half
m^2\F(iy)^2\right)\ =\nn &&-i\int\dd
q\,q^0\left(\hat{a_y}(q)a_y(q)+\half\right)\ =\ -i\int\dd
q\,q^0(n+\half)\ .\qquad \eel{Hamy} Clearly, with the hermiticity
condition \eqn{aydagger}, the Hamiltonian became purely imaginary,
as in Section \ref{classfield.sec}. Also, the zero point
fluctuations still seem to be there. However, we have not yet
addressed the operator ordering. Let us take a closer look at the
way individual creation and annihilation operators transform. We
now need to set \(m=0,\ p^0=|p|,\ q^0=|q|\). In order to compare
the creation and annihilation operators in real space-time with
those in imaginary space-time, substitute Eqs.~\eqn{fieldycreate}
and \eqn{conjycreate} into \eqn{annihilfield}, and the converse,
to obtain \be \fl\quad a(p)&=&\int\int{\dd x\dd q\over
2\pi\sqrt{4p^0q^0}}\bigg\{(p^0-iq^0)\,a_y(q)\,e^{(q-ip)x}+(p^0+iq^0)\,\hat{a_y}(q)\,e^{(-q-ip)x}\bigg\}\
,\qquad \crl{aay}\fl\quad a_y(q)&=&\int\int{\dd y\dd
p\over2\pi\sqrt{4p^0q^0}}\bigg\{(q^0+ip^0)\,a(p)\,e^{(-iq-p)y}+(q^0-ip^0)\,a^\dagger(p)\,e^{(-iq+p)y}\bigg\}
\ . \qquad\eel{aya}

The difficulty with these expressions is the fact that the \(x\)- and the \(y\)-integrals diverge. We now propose the following procedure. Let us
limit ourselves to the case that, in Eqs.~\eqn{annihilyfield} and \eqn{createyfield}, the \(y\)-integration is over a finite box only: \(|y|<L\), in
which case \(a_y(q)\sqrt{2q^0}\) will be an entire analytic function of \(q\). Then, in Eq.~\eqn{aay}, we can first shift the integration contour in
complex \(q\)-space by an amount \(ip\) up or down, and subsequently rotate the \(x\)-integration contour to obtain convergence. Now the square roots
occurring explicitly in Eqs.~\eqn{aay} and \eqn{aya} are merely the consequence of a choice of normalization, and could be avoided, but the root in
the definition of \(p^0\) and \(q^0\) are much more problematic. In principle we could take any of the two branches of the roots. However, in our
transformation procedure we actually \emph{choose} \(q^0=-ip^0\) and the second parts of Eqs.~\eqn{aay} and \eqn{aya} simply cancel out. Note that,
had we taken the other sign, i.e. \(q^0=ip^0\), this would have affected the expression for \(\F(iy,\,i\t)\) such, that we would still end up with
the same final result. In general, the \(x\)-integration yields a delta function constraining \(q\) to be \(\pm ip\), but \(q^0\) is chosen to be on
the branch \(-ip^0\), in both terms of this equation (\(q^0\) normally does not change sign if \(q\) does). Thus, we get, from Eqs.~\eqn{aay}
and \eqn{aya}, respectively, \be a(p)&=&i^{1/2}\,a_y(q)\ ,\qquad q=ip\,,\quad q^0=ip^0\ ,\\
a_y(q)&=&i^{-1/2}\,a(p)\ ,\qquad p=-iq\,,\quad p^0=-iq^0\
,\eel{apayq} so that \(a(p)\) and \(a_y(q)\) are analytic
continuations of one another. Similarly, \be\fl\quad
a^\dagger(p)=i^{1/2}\,\hat{a_y}(q)\ ,\qquad \hat{a_y}(q)=
i^{-1/2}\,a^\dagger(p)\ ,\qquad p=-iq\,,\quad p^0=-iq^0\
.\eel{ahatdagger} There is no Bogolyubov mixing between \(a\) and
\(a^\dagger\). Note that these expressions agree with the
transformation law of the Hamiltonian \eqn{Hamy}.

Now that we have a precisely defined transformation law for the creation and annihilation operators, we can find out how the states transform. The
vacuum state \(|0\ket\) is defined to be the state upon which all annihilation operators vanish. We now see that this state is invariant under all
our transformations. Indeed, because there is no Bogolyubov mixing, all \(N\) particle states transform into \(N\) particle states, with \(N\) being
invariant. The vacuum is invariant because 1) unlike the case of the harmonic oscillator, Section~\ref{harmonic.sec}, creation operators transform
into creation operators, and annihilation operators into annihilation operators, and because 2) the vacuum is translation invariant.

The Hamiltonian is transformed into \(-i\) times the Hamiltonian (in the case \(D=2\)); the energy density \(T_{00}\) into \(-T_{00}\), and since the
vacuum is the only state that is invariant, it must have \(T_{00}=0\) and it must be the only state with this property.

\newsec{Pure Maxwell Fields}

This can now easily be extended to include the Maxwell action as well. In flat spacetime: \be S = -\int
d^3x\,\frac{1}{4}F_{\m\n}(x)F^{\m\n}(x),\quad\quad F_{\m\n}=\pa_\m A_\n - \pa_\n A_\m. \eel{actMaxx} The action is invariant under gauge
transformations of the form \be A_\m(x)\ra A_\m(x) + \pa_\m\x(x).\eel{gaugetrafo} Making use of this freedom, we impose the Lorentz condition
\(\pa_\m A^\m=0\), such that the equation of motion \(\pa_\m F^{\m\n}=0\) becomes \(\Box A^\m =0\). As is well known, this does not completely fix
the gauge, since transformations like \eqn{gaugetrafo} are still possible, provided \(\Box\x=0\). This remaining gauge freedom can be used to set
\(\na\cdot \vec{A}=0\), denoted Coulomb gauge, which sacrifices manifest Lorentz invariance. The commutation relations are \be [E^i(x,t),\,A_j(x',t)]
= i\d_{ij}^{tr}(\vec{x}-\vec{x}'),\eel{commxmax} where \be E^k=\frac{\pa\LL}{\pa\dot{A}_k} = -\dot{A}^k -\frac{\pa A_0}{\pa x^k},\ee is the momentum
conjugate to \(A^k\), which we previously called \(\Pi\), but it is here just a component of the electric field. The transverse delta function is
defined as \be \d_{ij}^{tr}(\vec{x}-\vec{x}')\equiv\int\frac{d^3p}{(2\pi)^3}e^{i\vec{p}(\vec{x}-\vec{x}')} \left(\d_{ij} -
\frac{p_ip_j}{\vec{p}^{\;2}}\right),\eel{deltatrx} such that its divergence vanishes. In Coulomb gauge, \(\vec{A}\) satisfies the wave equation
\(\Box\vec{A}=0\), and we write \be \vec{A}(x,t)=\int{\dd^3 p\over(2\pi)^3\sqrt{2p^0}}\sum_{\l=1}^{2}\vec{\e}(p,\,\l)\Big(
a(p,\,\l)e^{i(px)}+a^\dag(p,\,\l) e^{-i(px)}\Big)\ ,\ee where \(\vec{\e}(p,\,\l)\) is the polarization vector of the gauge field, which satisfies
\(\vec{\e}\cdot\vec{p}=0\) from the Coulomb condition \(\na\cdot\vec{A}=0\). Moreover, the polarization vectors can be chosen to be orthogonal
\(\vec{\e}(p,\,\l)\cdot\vec{\e}(p,\,\l')=\d_{\l\l'}\) and satisfy a completeness relation
\be\sum_{\l}\e_m(p,\,\l)\e_n(p\,\l)=\left(\delta_{mn}-\frac{p_mp_n}{\vec{p}^{\;2}}\right).\eel{complpol} The commutator between the creation and
annihilation operators becomes \be [a(p,\,\l),\,a^\dag(p',\,\l)]=\d(\vec{p}-\vec{p}')\d_{\l\l'},\ee in which the polarization vectors cancel out due
to their completeness relation.

In complex space, the field \(A_\m\) thus transforms analogously to the scalar field, with the only addition that the polarization vectors
\(\vec{\e}_\m(p)\) will now become function of complex momentum \(\vec{q}\). However, since they do not satisfy a particular algebra, like the
creation and annihilation operators, they do not cause any additional difficulties. The commutation relations between the creation and annihilation
operators behave similarly as in the scalar field case, since the second term in the transverse delta function \eqn{deltatrx}, and the polarization
vector completeness relation \eqn{complpol}, is invariant when transforming to complex momentum.

Thus we find \be F_{\m\n}(x,t)F^{\m\n}(x,t)\ra -F_{\m\n}(iy,i\t)F^{\m\n}(iy,i\t),\ee and again $T_{00}$ flips sign, as the energy-momentum tensor
reads: \be T_{\m\n} = -F_{\m\a}F_{\n}^{\a} + \frac{1}{4}F_{\a\b}F^{\a\b}\h_{\m\n}.\ee In term of the \(E\) and \(B\) fields, which are given by
derivatives of \(A_\m\), \(E_i=F_{0i}\), \(B_k=\half\e_{ijk}F_{jk}\), we have: \be T_{00}=\half\left(E^2 + B^2\right)\ra - T_{00},\ee which indicates
that the electric and magnetic fields become imaginary. A source term \(J^\m A_\m\) can also be added to the action \eqn{actMaxx}, if one imposes
that \(J^\m\ra -J^\m\), in which case the Maxwell equations \(\pa_\m F^{\m\n}=J^\n\) are covariant.

Implementing gauge invariance in imaginary space is also straightforward. The Max\-well action and Max\-well equations are invariant under
\(A_\m(x,t)\ra A_\m(x,t) +\pa_\m\x(x,t))\). In complex spacetime this becomes \be A_\m(iy,i\t)\ra A_\m(iy,i\t) -i\pa_\m(y,\t)\x(iy,i\t)\ee and the
Lorentz condition \be\pa_\m(x,t) A^\m(x,t) =0\quad \ra\quad -i\pa(y,\t)A^\m(iy,i\t).\ee In Coulomb gauge the polarization vectors satisfy
\be\vec{\e}(q)\cdot\vec{q}=0\ ,\ee with imaginary momentum \(q\).

Unfortunately, the Maxwell field handled this way will not be easy to generalize to Yang-Mills fields. The Yang-Mills field has cubic and quartic
couplings, and these will transform with the wrong sign. One might consider forcing vector potentials to transform just like space-time derivatives,
but then the kinetic term emerges with the wrong sign. Alternatively, one could suspect that the gauge group, which is compact in real space, would
become non-compact in imaginary space, but this also leads to undesirable features.

\newsec{Relation with Boundary Conditions}

All particle states depend on boundary conditions, usually imposed on the real axis. One could therefore try to simply view the \(x\ra ix\) symmetry
as a one-to-one mapping of states with boundary conditions imposed on \(\pm x\ra\infty\) to states with boundary conditions imposed on imaginary axis
\(\pm ix\ra\infty\). At first sight, this mapping transforms positive energy particle states into negative energy particle states. The vacuum, not
having to obey boundary conditions would necessarily have zero energy. However, this turns out not to be sufficient.

Solutions to the Klein-Gordon equation, with boundary conditions
imposed on imaginary coordinates are of the form: \be
\fl\quad\F_\mathrm{im}(x,\,t)=\quad\int{\dd p\over\sqrt{2\pi\cdot
2p^0}}\left(a(p)e^{(px)}+\hat{a}(p)e^{-(px)}\right)\ ,&\quad&
p^0=\sqrt{p^2+m^2},\eel{bccompl} written with the subscript `im'
to remind us that this is the solution with boundary conditions on
the imaginary axis. With these boundary conditions, the field
explodes for real valued \(x\ra\pm\infty\), whereas for the usual
boundary conditions, imposed on the real axis, the field explodes
for \(ix\ra\pm\infty\). Note that for non-trivial \(a\) and
\(\hat{a}\), this field now has a non-zero complex part on the
real axis, if one insists that the second term is the Hermitian
conjugate of the first, as is usually the case. This is a
necessary consequence of this setup. However, we insist on writing
\(\hat{a}=a^\dag\) and, returning to three spatial dimensions, we
write for \(\F_\mathrm{im}(x,\,t)\) and
\(\Pi_\mathrm{im}(x,\,t)\): \be &\F_\mathrm{im}(\vec{x},t)& =
\int\frac{d^3p}{(2\pi)^3}\frac{1}{\sqrt{2p^0}}\left(a_pe^{(px)}
+ a^\dag_{p}e^{-(px)}\right),\nonumber\\
\dot{\F}_\mathrm{im}(\vec{x},t) = &\Pi_\mathrm{im}(\vec{x},t)& =
\int\frac{d^3p}{(2\pi)^3}(-){\sqrt{\frac{p^0}{2}}}\left(a_pe^{(
px)} - a^{\dag}_pe^{-(px)}\right),\nonumber\\
&&p^0 = \vphantom{\Big[}\sqrt{\vec p^{\;2}+m^2}\ ,\qquad(px)\deff
\vec p\cdot\vec x-p^0t,\ee and impose the normal commutation
relations between \(a\) and \(a^\dag\): \be
[a_p,a_{p'}^{\dag}]=(2\pi)^3\delta^{(3)}(\vec{p}-\vec{p}').\eel{commui}
Using Eqn. \eqn{commui}, the commutator between \(\F_\mathrm{im}\)
and \(\Pi_\mathrm{im}\) at equal times, becomes: \be
[\F_\mathrm{im}(\vec{x}),\,\Pi_\mathrm{im}(\vec{x})] =
\delta^{(3)}(\vec{x}-\vec{x}'),\ee which differs by a factor of
\(i\) from the usual relation, and by a minus sign, compared to
Eqn. \eqn{complcomm}. The energy-momentum tensor is given by \be
T_{\m\n\,\mathrm{im}}=\pa_\m\F_\mathrm{im}\pa_\n\F_\mathrm{im} -
\half \h_{\m\n}\pa^k\F_\mathrm{im}\pa_k\F_\mathrm{im}, \ee and
thus indeed changes sign, as long as one considers only those
contributions to a Hamiltonian that contain products of \(a\) and
\(a^\dagger\): \be H_\mathrm{im}^\mathrm{diag}=
\int\frac{d^3p}{(2\pi)^3}p^0\left(-a_{p}^{\dag}a_p
-\frac{1}{2}[a_p,\,a_{p}^{\dag}]\right)=-H.\ee However, the
remaining parts give a contribution that is rapidly diverging on
the imaginary axis \be T_{\m\n\,\mathrm{im}}^\mathrm{non-diag}=
a^2e^{2(px)} + (a^{\dag})^2e^{-2(px)},\ee but which blows up for
\(\pm x\ra\infty\). Note that when calculating vacuum expectation
values, these terms give no contribution.

To summarize, one can only construct such a symmetry, changing boundary conditions from real to imaginary coordinates, in a very small box. This was
to be expected, since we are comparing hyperbolic functions with their ordinary counterparts, \(\sinh(x)\) vs. \(\sin(x)\), and they are only
identical functions in a small neighborhood around the origin.

\newsec{Conclusions}

It is natural to ascribe the extremely tiny value of the cosmological constant to some symmetry. Until now, the only symmetry that showed promises in
this respect has been supersymmetry. It is difficult, however, to understand how it can be that supersymmetry is obviously strongly broken by all
matter representations, whereas nevertheless the vacuum state should respect it completely. This symmetry requires the vacuum fluctuations of bosonic
fields to cancel very precisely against those of the fermionic field, and it is hard to see how this can happen when fermionic and bosonic fields
have such dissimilar spectra.

The symmetry proposed in this paper is different. It is suspected that the field equations themselves have a larger symmetry than the boundary
conditions for the solutions. It is the boundary conditions, and the hermiticity conditions on the fields, that force all physical states to have
positive energies. If we drop these conditions, we get as many negative energy as positive energy states, and indeed, there may be a symmetry
relating positive energy with negative energy. This is the most promising beginning of an argument why the vacuum state must have strictly vanishing
gravitational energy.

The fact that the symmetry must relate real to imaginary coordinates is suggested by the fact that De Sitter and Anti-De Sitter space are related by
analytic continuation, and that their cosmological constants have opposite sign.

Unfortunately, it is hard to see how this kind of symmetry could be realized in the known interaction types seen in the sub-atomic particles. At
first sight, all mass terms are forbidden. However, we could observe that all masses in the Standard Model are due to interactions, and it could be
that fields with positive mass squared are related to tachyonic fields by our symmetry. The one scalar field in the Standard Model is the Higgs
field. Its self interaction is described by a potential \(V_1(\F)=\half\l(\F^\dagger\F-F^2)^2\), and it is strongly suspected that the parameter
\(\l\) is unnaturally small. Our symmetry would relate it to another scalar field with opposite potential: \(V_2(\F_2)=-V_1(\F_2)\). Such a field
would have no vacuum expectation value, and, according to perturbation theory, a mass that is the Higgs mass divided by \(\sqrt 2\). Although
explicit predictions would be premature, this does suggest that a theory of this kind could make testable predictions, and it is worth-while to
search for scalar fields that do not contribute to the Higgs mechanism at LHC, having a mass somewhat smaller than the Higgs mass. We are hesitant
with this particular prediction because the negative sign in its self interaction potential could lead to unlikely instabilities, to be taken care of
by non-perturbative radiative corrections.

The symmetry we studied in this paper would set the vacuum energy
to zero and has therefore the potential to explain a vanishing
cosmological constant. In light of the recent discoveries that the
universe appears to be accelerating
\cite{Riess1998,Perlmutter1998,Riess2004}, one could consider a
slight breaking of this symmetry. This is a non-trivial task that
we will have to postpone to further work. Note however, that our
proposal would only nullify exact vacuum energy with equation of
state \(w=-1\). Explaining the acceleration of the universe with
some dark energy component other than a cosmological constant,
quintessence for example, therefore is not ruled out within this
framework.

The considerations reported about in this paper will only become
significant if, besides Maxwell fields, we can also handle
Yang-Mills fields, fermions, and more complicated interactions. As
stated, Yang-Mills fields appear to lead to difficulties.
Fermions, satisfying the linear Dirac equation, can be handled in
this formalism. Just as is the case for scalar fields, one finds
that mass terms are forbidden for fermions, but we postpone
further details to future work. Radiative corrections and
renormalization group effects will have to be considered. To stay
in line with our earlier paper, we still consider arguments of
this nature to explain the tiny value of the cosmological constant
unlikely to be completely successful, but every alley must be
explored, and this is one of them.

\newsec{References}
\bibliography{xnaarix}

\providecommand{\href}[2]{#2}\begingroup\raggedright\begin{thebibliography}{10}

\bibitem{Nobbenhuis:2004wn}
S.~Nobbenhuis, ``Categorizing different approaches to the cosmological constant
  problem,'' {\em Found. Phys.} {\bf 36} (2006)
\href{http://www.arXiv.org/abs/gr-qc/0411093}{{\tt gr-qc/0411093}}.
%%CITATION = GR-QC 0411093;%%.

\bibitem{Kaplan:2005rr}
D.~E. Kaplan and R.~Sundrum, ``A symmetry for the cosmological constant,''
\href{http://www.arXiv.org/abs/hep-th/0505265}{{\tt hep-th/0505265}}.
%%CITATION = HEP-TH 0505265;%%.

\bibitem{Bender:2005pf}
C.~M. Bender, ``Non-Hermitian quantum field theory,'' {\em Int. J. Mod. Phys.}
  {\bf A20} (2005)
4646--4652.
%%CITATION = IMPAE,A20,4646;%%.

\bibitem{Bender:2005hf}
C.~M. Bender, H.~F. Jones, and R.~J. Rivers, ``Dual PT-symmetric quantum field
  theories,'' {\em Phys. Lett.} {\bf B625} (2005) 333--340,
\href{http://www.arXiv.org/abs/hep-th/0508105}{{\tt hep-th/0508105}}.
%%CITATION = HEP-TH 0508105;%%.

\bibitem{Bender:1998ke}
C.~M. Bender and S.~Boettcher, ``Real Spectra in Non-Hermitian Hamiltonians
  Having PT Symmetry,'' {\em Phys. Rev. Lett.} {\bf 80} (1998) 5243--5246,
\href{http://www.arXiv.org/abs/physics/9712001}{{\tt physics/9712001}}.
%%CITATION = PHYSICS 9712001;%%.

\bibitem{Bender2005}
A.~Z., C.~M. Bender, and M.~Berry, ``Reflectionless Potentials and PT
  Symmetry,'' \href{http://www.arXiv.org/abs/quant-ph/0508117}{{\tt
  quant-ph/0508117}}.

\bibitem{Bonelli:2000tz}
G.~Bonelli and A.~M. Boyarsky, ``Six dimensional topological gravity and the
  cosmological constant problem,'' {\em Phys. Lett.} {\bf B490} (2000)
  147--153,
\href{http://www.arXiv.org/abs/hep-th/0004058}{{\tt hep-th/0004058}}.
%%CITATION = HEP-TH 0004058;%%.

\bibitem{Erdem2004}
R.~Erdem, ``A symmetry for vanishing cosmological constant in an extra
  dimensional toy model,'' {\em Phys. Lett.} {\bf B621} (2005) 11--17,
\href{http://www.arXiv.org/abs/hep-th/0410063}{{\tt hep-th/0410063}}.
%%CITATION = HEP-TH 0410063;%%.

\bibitem{Jackiw}
 Private communication.

\bibitem{Riess1998}
{\bf Supernova Search Team} Collaboration, A.~G. Riess {\em et al.},
  ``Observational evidence from supernovae for an accelerating universe and a
  cosmological constant,'' {\em Astron. J.} {\bf 116} (1998) 1009--1038,
\href{http://www.arXiv.org/abs/astro-ph/9805201}{{\tt astro-ph/9805201}}.
%%CITATION = ASTRO-PH 9805201;%%.

\bibitem{Perlmutter1998}
{\bf Supernova Cosmology Project} Collaboration, S.~Perlmutter {\em et al.},
  ``Measurements of omega and lambda from 42 high-redshift supernovae,'' {\em
  Astrophys. J.} {\bf 517} (1999) 565--586,
\href{http://www.arXiv.org/abs/astro-ph/9812133}{{\tt astro-ph/9812133}}.
%%CITATION = ASTRO-PH 9812133;%%.

\bibitem{Riess2004}
{\bf Supernova Search Team} Collaboration, A.~G. Riess {\em et al.}, ``Type Ia
  supernova discoveries at z>1 from the Hubble space telescope: evidence for
  past deceleration and constraints on dark energy evolution,'' {\em Astrophys.
  J.} {\bf 607} (2004) 665--687,
\href{http://www.arXiv.org/abs/astro-ph/0402512}{{\tt astro-ph/0402512}}.
%%CITATION = ASTRO-PH 0402512;%%.

\end{thebibliography}\endgroup
\bibliographystyle{utcaps}

\end{document}